\documentclass[amsmath,amssymb,prc,final,superscriptaddress,showpacs,twocolumn]{revtex4} 

\usepackage{bm,hyphenat,xspace,draftcopy}
\usepackage{graphicx,epsfig}

\newcommand{\scl}{0.63}

\newcommand{\Eq}{Eq.}

\newcommand{\Fig}{Fig.}
\newcommand{\Figs}{Figs.}
\newcommand{\Ref}{Ref.}
\newcommand{\Refs}{Refs.}
\newcommand{\Sect}{Sec.}
\newcommand {\mbf}[1]{{\mathbf{#1}}}

\newcommand {\Kpl}{\mbf{K}_{+}}

\newcommand{\fm}{\;\mathrm{fm}}
\newcommand{\cm}{\mathrm{c\!\:\!.m\!\:\!.}}
\newcommand{\He}{{}^3\mathrm{He}}
\newcommand{\Hh}{{}^3\mathrm{H}}
\newcommand{\zr}{z_R^{-\frac12}}
\newcommand{\ZR}{\mathcal{Z}_{R}^{-\frac12}}
\newcommand{\zR}{\mathcal{Z}}

\begin{document}

\title {Momentum-space description of three-nucleon breakup reactions including
the Coulomb interaction}

\author{A.~Deltuva} 
\email{deltuva@cii.fc.ul.pt}
\thanks{on leave from Institute of Theoretical Physics and Astronomy,
Vilnius University, Vilnius 2600, Lithuania}
\affiliation{Centro de F\'{\i}sica Nuclear da Universidade de Lisboa, 
P-1649-003 Lisboa, Portugal }

\author{A.~C.~Fonseca} 
\affiliation{Centro de F\'{\i}sica Nuclear da Universidade de Lisboa, 
P-1649-003 Lisboa, Portugal }

\author{P.~U.~Sauer}
\affiliation{Institut f\"ur Theoretische Physik,  Universit\"at Hannover,
  D-30167 Hannover, Germany}
\received{August 4, 2005}


\begin{abstract}
The Coulomb interaction between the two protons is included in the calculation
of proton-deuteron breakup  
and of three-body electromagnetic disintegration of ${}^3\mathrm{He}$.
The hadron dynamics is based on the purely nucleonic
charge-dependent (CD) Bonn potential and its realistic extension
CD Bonn + $\Delta$ to a coupled-channel two-baryon potential,
allowing for single virtual $\Delta$-isobar excitation.
Calculations are done using integral equations in momentum space.
The screening and renormalization approach is employed for including the
Coulomb interaction.
Convergence of the procedure is found at moderate screening radii.
The reliability of the method is demonstrated.
The Coulomb effect on breakup observables is  seen at all energies
in particular kinematic regimes.
\end{abstract}

 \maketitle

\section{Introduction \label{sec:intro}}

The inclusion of the Coulomb interaction in the description of the
three-nucleon continuum is one of the most challenging tasks in 
theoretical few-body nuclear physics~\cite{alt:02a}. 
Whereas it has already been solved for elastic
proton-deuteron $(pd)$ scattering with realistic hadronic interactions
using various procedures 
\cite{alt:02a,kievsky:01a,chen:01a,ishikawa:03a,deltuva:05a},
there are only very few attempts \cite{alt:94a,kievsky:97a,suslov:04a}
to calculate $pd$ breakup, and none of them uses a complete treatment
of the Coulomb interaction and realistic hadronic potentials 
allowing for a stringent comparison with the experimental data.

Recently in \Ref~\cite{deltuva:05a} we included the Coulomb interaction 
between the protons in the description of three-nucleon reactions
with two-body initial and final states.
The description is based on the Alt-Grassberger-Sandhas (AGS) 
equation~\cite{alt:67a} in momentum space. The Coulomb potential
is screened and the resulting scattering amplitudes are corrected  
by the renormalization technique of \Refs~\cite{taylor:74a,alt:78a} 
to recover the unscreened limit.
The treatment is applicable to any two-nucleon potential without separable
expansion. 
Reference \cite{deltuva:05a} and this paper use the purely nucleonic
charge-dependent (CD) Bonn potential \cite{machleidt:01a} 
and its  coupled-channel extension CD Bonn + $\Delta$ \cite{deltuva:03c}, 
allowing for a single  virtual $\Delta$-isobar excitation and fitted 
to the experimental data with the same degree of accuracy as CD Bonn itself. 
In the three-nucleon system the $\Delta$ isobar mediates an effective 
three-nucleon force and effective two- and three-nucleon currents,
both consistent with the underlying two-nucleon force.
The treatment of \Ref~\cite{deltuva:05a} is technically highly successful,
but still limited to the description of proton-deuteron $(pd)$ elastic
scattering and of electromagnetic (e.m.) reactions involving $\He$ with
$pd$ initial or final states only. This paper extends the treatment
of Coulomb to breakup in $pd$ scattering and to e.m. three-body
disintegration of $\He$. In that extension we follow the ideas of 
\Refs~\cite{taylor:74a,alt:78a,alt:94a}, but avoid 
approximations on the hadronic potential and in the treatment of 
screened Coulomb. Thus, our three-particle equations,
including the screened Coulomb potential, are completely different
from the quasiparticle equations solved in \Ref~\cite{alt:94a}
where the two-nucleon screened Coulomb transition matrix is
approximated by the screened Coulomb potential.
In \Ref~\cite{deltuva:05c} we presented for the first time a limited
set of results for $pd$ breakup using the same technical developments
we explain here in greater detail.

We have to recall that the screened Coulomb potential $w_R$ we work with
is particular. It is screened around the separation $r=R$ 
between two charged baryons and in configuration space is given by
\begin{gather} \label{eq:wr}
w_R(r) = w(r) \; e^{-(r/R)^n},
\end{gather}
with the true Coulomb potential $w(r)=\alpha_e/r$, 
$\alpha_e$ being the fine structure constant and
$n$ controlling the smoothness of the screening. 
We prefer to work with a sharper screening than the Yukawa screening
$(n=1)$ of \Ref~\cite{alt:94a}. We want to ensure that the 
screened Coulomb potential $w_R$ approximates well the true Coulomb one
$w$ for distances $r<R$  and simultaneously vanishes rapidly for $r>R$, 
providing a comparatively fast convergence of the partial-wave expansion.
 In contrast, the sharp cutoff  $(n \to \infty)$
yields an unpleasant oscillatory behavior in the momentum-space representation,
leading to convergence problems. 
We find the values $3 \le n \le 6$ to provide a sufficiently smooth, 
but at the same time a sufficiently rapid screening around $r=R$
like in \Ref~\cite{deltuva:05a}; 
$n=4$ is our choice for the results of this paper.
The screening radius $R$ is chosen much larger than the range of the
strong interaction which is of the order of the pion wavelength
$\hbar/m_\pi c \approx 1.4\fm$. Nevertheless, 
the screened Coulomb potential $w_R$ is of short range in the sense
of scattering theory. Standard scattering theory is therefore applicable.
A reliable technique \cite{deltuva:03a} for solving the 
AGS equation~\cite{alt:67a} with short-range interactions 
is extended in \Ref~\cite{deltuva:05a} to include the screened Coulomb 
potential between the charged baryons.
However, the partial-wave expansion of the pair interaction requires
much higher angular momenta than the one of the strong two-nucleon potential
alone.

The screening radius $R$ will always remain very small compared with
nuclear screening distances of atomic scale, i.e., 
$10^5\fm$. Thus, the employed screened Coulomb potential $w_R$ 
is unable to simulate properly the physics of nuclear screening 
and, even more, all features of the true Coulomb potential. 
Thus, the approximate breakup calculations with screened Coulomb 
$w_R$ have to be corrected for their shortcomings in a controlled way.
References~\cite{taylor:74a,alt:78a} 
give the prescription for the correction procedure which we follow here
for breakup as we did previously for elastic scattering,
and that involves the renormalization of the on-shell amplitudes
in order to get the proper unscreened Coulomb limit.
 After the indicated corrections, the predictions for breakup observables
have to show independence from the choice
of the screening radius $R$, provided it is chosen sufficiently large.
That convergence will be the internal criterion for the reliability
of our Coulomb treatment.

Configuration space treatments of Coulomb \cite{kievsky:97a,suslov:04a}
may provide a viable alternative to the integral equation approach in
momentum space on which this paper is based. 
References \cite{kievsky:97a,suslov:04a} have provided first results for 
$pd$ breakup, but they still involve approximations in the treatment of
Coulomb and the employed hadronic dynamics is not  realistic. 
Thus, a benchmark comparison between our breakup results and corresponding
configuration space results is, in contrast to $pd$ elastic 
scattering \cite{deltuva:05b}, not possible yet. With respect to the
reliability of our Coulomb treatment for breakup, we rely solely
on our internal criterion, i.e., the convergence of breakup observables
with the screening radius $R$; however, that criterion was absolutely
reliable for $pd$ elastic scattering and related e.m. reactions.

Section~\ref{sec:th} develops the technical apparatus underlying the
calculations. 
Section~\ref{sec:res} presents some characteristic effects of Coulomb
in three-nucleon breakup reactions.
Section~\ref{sec:concl} gives our conclusions.

\section{Treatment of Coulomb interaction between protons in breakup
 \label{sec:th}}

This section carries over the treatment of the Coulomb interaction
given in \Ref~\cite{deltuva:05a} for $pd$ elastic scattering and 
corresponding e.m. reactions, to $pd$ breakup and to 
e.m. three-body disintegration of $\He$.
It establishes a theoretical procedure leading to a calculational scheme.
The discussions of hadronic and e.m. reactions are done separately.

\subsection{Theoretical framework for the description of proton-deuteron
breakup with Coulomb \label{sec:thpdb}}

This section focuses on $pd$ breakup. However, the transition matrices
for elastic scattering and breakup are so closely connected that certain
relations between scattering operators already developed in 
\Ref~\cite{deltuva:05a} have to be recalled to make this paper selfcontained.

Each pair of nucleons $(\beta \gamma)$ interacts through the strong 
coupled-channel potential $v_\alpha$ and the Coulomb potential $w_\alpha$. 
We assume that $w_\alpha$ acts formally between all pairs $(\beta \gamma)$
of particles, but it is nonzero only for states with two-charged baryons,
i.e., $pp$ and $p\Delta^+$ states. 
We introduce the full resolvent $G^{(R)}(Z)$ for the auxiliary situation
in which the Coulomb potential $w_\alpha$ is screened with a
screening radius $R$, $w_\alpha$ being replaced by $w_{\alpha R}$, 
\begin{gather} \label{eq:GR1}
G^{(R)}(Z) = (Z - H_0 - \sum_\sigma v_\sigma - \sum_\sigma w_{\sigma R})^{-1},
\end{gather}
where $H_0$ is the three-particle kinetic energy operator. The full resolvent
yields the full $pd$ scattering state when acting on the initial
channel state $ |\phi_\alpha (\mbf{q}_i) \nu_{\alpha_i} \rangle $ of
relative $pd$ momentum $\mbf{q}_i$, energy $E_\alpha(q_i)$ 
and additional discrete quantum numbers $\nu_{\alpha_i}$ and  taking the
appropriate limit $Z = E_\alpha(q_i) + i0$. 
The full $pd$ scattering state has, above breakup threshold, components
corresponding to the final breakup channel states 
$ |\phi_0 (\mbf{p}_f \mbf{q}_f) \nu_{0_f} \rangle $, $\, \mbf{p}_f$ and 
$\mbf{q}_f$ being three-nucleon Jacobi momenta,
$E_0 (p_f q_f)$ its energy, and $\nu_{0_f}$
additional discrete quantum numbers.
The full resolvent therefore also yields the desired $S$ matrix for breakup. 
The full resolvent $G^{(R)}(Z)$ depends on the screening radius $R$ 
for the Coulomb potential and that dependence is 
notationally indicated; the same will be done for operators related
to $G^{(R)}(Z)$. Following standard 
AGS notation~\cite{alt:67a} of three-particle scattering,
the full resolvent $G^{(R)}(Z)$ may be decomposed
into channel resolvents and free resolvent
\begin{subequations} \label{eq:GRa}
\begin{align} 
 G^{(R)}_\alpha (Z) = {} & (Z - H_0 - v_\alpha - w_{\alpha R})^{-1}, \\
G_0 (Z) = {} & (Z - H_0)^{-1},
\end{align}
\end{subequations}
together with the full multichannel three-particle transition matrices 
$U^{(R)}_{\beta \alpha}(Z)$ for elastic scattering and
$U^{(R)}_{0 \alpha}(Z)$ for breakup according to 
\begin{subequations} \label{eq:GU} 
  \begin{align} \label{eq:GUa}
    G^{(R)}(Z) = {} & \delta_{\beta \alpha}  G^{(R)}_\alpha (Z) +
    G^{(R)}_\beta (Z)  U^{(R)}_{\beta \alpha}(Z)   G^{(R)}_\alpha (Z), \\
    \label{eq:GU0}
    G^{(R)}(Z) = {} & G_0(Z) U^{(R)}_{0\alpha}(Z) G^{(R)}_\alpha (Z).
  \end{align}
\end{subequations}
The full multichannel transition matrices satisfy the 
AGS equations~\cite{alt:67a}
\begin{subequations}\label{eq:UbaT}
  \begin{align} \label{eq:Uba}
     U^{(R)}_{\beta \alpha}(Z) = {} & \bar{\delta}_{\beta \alpha} G_0^{-1}(Z)
     + \sum_{\sigma} \bar{\delta}_{\beta \sigma} T^{(R)}_\sigma (Z) G_0(Z) 
     U^{(R)}_{\sigma \alpha}(Z), \\
     U^{(R)}_{0 \alpha}(Z) = {} & G_0^{-1}(Z)
     + \sum_{\sigma}  T^{(R)}_\sigma (Z) G_0(Z) U^{(R)}_{\sigma \alpha}(Z), 
  \end{align}
with  $\bar{\delta}_{\beta \alpha} = 1 - {\delta}_{\beta \alpha}$;
the two-particle transition matrix $T^{(R)}_\alpha (Z)$
is derived from the full channel interaction $v_\alpha + w_{\alpha R}$
including screened Coulomb, i.e.,
 \begin{align} \label{eq:TR}
   T^{(R)}_\alpha (Z) = {}& (v_\alpha + w_{\alpha R}) + 
     (v_\alpha + w_{\alpha R})  G_0(Z) T^{(R)}_\alpha (Z).
  \end{align}
\end{subequations}

In $pd$ elastic scattering, an alternative decomposition of the full 
resolvent is found conceptually more revealing.
Instead of correlating the plane-wave channel state
$ |\phi_\alpha (\mbf{q}) \nu_\alpha \rangle $ in a single step to the full
scattering state by $G^{(R)}(Z)$, it may be correlated first to a screened
Coulomb state of proton and deuteron by the screened Coulomb potential
$W^{\cm}_{\alpha R}$ between a proton and the center of mass (c.m.) 
of the remaining neutron-proton $(np)$ pair in channel $\alpha$ through
\begin{subequations}
  \begin{align}
    G_{\alpha R}(Z) = {}& 
    (Z - H_0 - v_\alpha - w_{\alpha R} - W^{\cm}_{\alpha R})^{-1}, \\
    G_{\alpha R}(Z) = {}& G^{(R)}_{\alpha}(Z) + 
    G^{(R)}_{\alpha}(Z) T^{\cm}_{\alpha R}(Z) G^{(R)}_{\alpha}(Z), \\
T^{\cm}_{\alpha R} (Z) = {} & W^{\cm}_{\alpha R} + 
W^{\cm}_{\alpha R} G^{(R)}_{\alpha} (Z) T^{\cm}_{\alpha R} (Z),
  \end{align}
\end{subequations}
where, in each channel $\alpha$,  $w_{\alpha R}$ and $W^{\cm}_{\alpha R}$ 
are never simultaneously present: When $\alpha$ corresponds to a $pp$ pair,
$w_{\alpha R}$ is present and $W^{\cm}_{\alpha R} = 0$;
when $\alpha$ denotes an $np$ pair, $w_{\alpha R} = 0$
and $W^{\cm}_{\alpha R}$ is present. The same Coulomb correlation 
is done explicitly in both initial and final states.
Thus, the full resolvent can be decomposed, in alternative to 
\Eq~\eqref{eq:GUa}, as
\begin{gather}
  G^{(R)}(Z) =  \delta_{\beta \alpha}   G_{\alpha R}(Z) +
  G_{\beta R}(Z) \tilde{U}^{(R)}_{\beta\alpha}(Z) G_{\alpha R}(Z),
\end{gather}
yielding a new form for the full multichannel transition matrix
\begin{subequations}
\begin{gather} \label{eq:U-T}
  \begin{split}
    U^{(R)}_{\beta \alpha}(Z) = {} & 
    \delta_{\beta\alpha} T^{\cm}_{\alpha R}(Z)
    +  [1 + T^{\cm}_{\beta R}(Z) G^{(R)}_{\beta}(Z)] \\ & \times
    \tilde{U}^{(R)}_{\beta\alpha}(Z)  
	  [1 + G^{(R)}_{\alpha}(Z) T^{\cm}_{\alpha R}(Z)].
  \end{split}
\end{gather}
The reduced operator $ \tilde{U}^{(R)}_{\beta\alpha}(Z)$
may be calculated through the integral equation
\begin{gather}
  \begin{split}
 \tilde{U}^{(R)}_{\beta\alpha}(Z) = {} &
\bar{\delta}_{\beta \alpha} [G_{\alpha R}^{-1}(Z) + v_{\alpha}] +
{\delta}_{\beta \alpha} \mathcal{W}_{\alpha R}  \\ &
+ \sum_\sigma (\bar{\delta}_{\beta \sigma} v_\sigma +
{\delta}_{\beta \sigma} \mathcal{W}_{\beta R})
G_{\sigma R}(Z) \tilde{U}^{(R)}_{\sigma\alpha}(Z),
\label{eq:tU}
\end{split}
\end{gather}
\end{subequations}
which is driven by the strong potential $v_\alpha$  and the  potential 
of three-nucleon nature
$\mathcal{W}_{\alpha R} = \sum_{\sigma} 
( \bar{\delta}_{\alpha \sigma} w_{\sigma R} - 
\delta_{\alpha\sigma} W^{\cm}_{\sigma R} ) $.
This potential $\mathcal{W}_{\alpha R}$
accounts for the difference between the direct $pp$ Coulomb interaction
and the one that takes place between the proton and the c.m. of the remaining
bound as well as unbound  $np$ pair.
When calculated between on-shell screened $pd$ Coulomb states,
$\tilde{U}^{(R)}_{\beta\alpha}(Z)$ is of
short-range, even in the infinite $R$ limit.

In the same spirit, the final breakup state to be analyzed may not be
reached in a single step; instead it may be correlated first to a screened
Coulomb state between the charged particles whose
corresponding Coulomb resolvent keeps only the screened Coulomb
interaction, 
\begin{subequations}
  \begin{gather}
    G_R(Z) =  (Z - H_0 -  \sum_\sigma  w_{\sigma R})^{-1}.
  \end{gather}
In  the system of two protons and one neutron 
only the channel $\sigma = \rho$, corresponding to a correlated $pp$ pair,
contributes to $ G_R(Z)$,
\begin{align}
  G_R(Z) = {} & G_0(Z) + G_0(Z) T_{\rho R}(Z) G_0(Z), \\
  T_{\rho R}(Z) = {} & w_{\rho R} + w_{\rho R} G_0(Z) T_{\rho R}(Z),
\end{align}
\end{subequations}
making channel $\rho$ the most convenient choice for the
description of the final breakup state.
Thus, for the purpose of $pd$ breakup, a decomposition of the full
resolvent, alternative to \Eq~\eqref{eq:GU0} is
\begin{subequations}
  \begin{align}
    G^{(R)}(Z) = {} & 
    G_{R}(Z) \tilde{U}^{(R)}_{0\alpha}(Z) G_{\alpha R}(Z), \\ 
    G^{(R)}(Z) = {} &  G_{0}(Z)  [1 + T_{\rho R}(Z) G_{0}(Z)]
    \tilde{U}^{(R)}_{0\alpha}(Z) \nonumber \\ & \times
	  [1 + G^{(R)}_{\alpha}(Z) T^{\cm}_{\alpha R}(Z)] G^{(R)}_{\alpha}(Z),
	  \label{eq:GRtU0}
  \end{align}
\end{subequations}
where the full breakup transition matrix may be written as
\begin{subequations}
  \begin{gather}\label{eq:U0t}
    \begin{split}
      U^{(R)}_{0\alpha}(Z) = {} & [1 + T_{\rho R}(Z) G_{0}(Z)]
      \tilde{U}^{(R)}_{0\alpha}(Z) \\ & \times
	    [1 + G^{(R)}_{\alpha}(Z) T^{\cm}_{\alpha R}(Z)],
    \end{split}
  \end{gather}
The reduced operator $ \tilde{U}^{(R)}_{0\alpha}(Z)$ may be calculated
through quadrature
\begin{gather} \label{eq:tU0}
    \tilde{U}^{(R)}_{0\alpha}(Z) = 
    G_{\alpha R}^{-1}(Z) + v_{\alpha}  
    + \sum_\sigma  v_\sigma G_{\sigma R}(Z) \tilde{U}^{(R)}_{\sigma\alpha}(Z)
\end{gather}
\end{subequations}
from the correspondingly reduced operator $\tilde{U}^{(R)}_{\beta\alpha}(Z)$
of elastic scattering.
In the form \eqref{eq:U0t} for the full breakup transition matrix the 
external distortions due to screened Coulomb in the initial and final
states are made explicit. On-shell the reduced operator 
$\tilde{U}^{(R)}_{0\alpha}(Z)$ calculated between screened Coulomb 
distorted initial and final states is of finite range and 
has two contributions with slightly different range properties: 

(a) The contribution $G_{\alpha R}^{-1}(Z) + v_{\alpha}$, when calculated
on-shell between initial $pd$ and final three-nucleon states, becomes
the three-nucleon potential $\mathcal{W}_{\alpha R}$ and is the
longest-range part of breakup, since the $np$ pair is correlated by the
hadronic interaction only in the initial $pd$ state.
The corresponding contribution in \Ref~\cite{alt:94a} is 
called the pure Coulomb breakup term.

(b) The remaining part
$\sum_\sigma  v_\sigma G_{\sigma R}(Z) \tilde{U}^{(R)}_{\sigma\alpha}(Z)$
is of shorter range, comparable to the
one of the reduced operator $ \tilde{U}^{(R)}_{\beta\alpha}(Z)$
for elastic $pd$ scattering.

In the full breakup operator $U^{(R)}_{0 \alpha}(Z)$
the external distortions show up in screened
Coulomb waves generated by $[1 + G^{(R)}_{\alpha}(Z) T^{\cm}_{\alpha R}(Z)]$
in the initial state and by $[1 + T_{\rho R}(Z) G_{0}(Z)]$ in the final
state; both wave functions do not have proper limits as $R \to \infty$.
Therefore $U^{(R)}_{0\alpha}(Z)$ has to get renormalized as the 
corresponding amplitude for $pd$ elastic scattering
\cite{deltuva:05a,alt:78a}, in order to obtain the results
appropriate for the unscreened Coulomb limit.
According to \Refs~\cite{alt:78a,alt:94a}, the full breakup transition
amplitude for initial and final states
$ |\phi_\alpha (\mbf{q}_i) \nu_{\alpha_i} \rangle $ and
$ |\phi_0 (\mbf{p}_f \mbf{q}_f) \nu_{0_f} \rangle $,
$E_\alpha(q_i)= E_0(p_f q_f)$, referring
to the strong potential $v_\alpha$ and the unscreened Coulomb potential 
$w_\alpha$, is obtained via the renormalization of the on-shell
breakup transition matrix $ U^{(R)}_{0 \alpha}(E_\alpha(q_i) + i0)$
in the infinite $R$ limit
  \begin{gather} \label{eq:UC1}
    \begin{split}
      \langle \phi_0 & (\mbf{p}_f \mbf{q}_f) \nu_{0_f} | U_{0 \alpha}
      |\phi_\alpha (\mbf{q}_i) \nu_{\alpha_i} \rangle  \\ = {}& 
      \lim_{R \to \infty} \{ \zr(p_f) 
      \langle \phi_0 (\mbf{p}_f \mbf{q}_f) \nu_{0_f} | \\ & \times
      U^{(R)}_{0 \alpha}(E_\alpha(q_i) + i0) 
      |\phi_\alpha (\mbf{q}_i) \nu_{\alpha_i} \rangle \ZR(q_i) \},
    \end{split}
  \end{gather}
where  $\zR_{R}(q_i)$ and $z_R(p_f)$ are $pd$ and $pp$ renormalization
factors defined below.

As in \Ref~\cite{deltuva:05a} we choose an isospin description
for the three baryons in which the nucleons are considered identical.
The two-baryon transition matrix $T^{(R)}_{\alpha}(Z)$ 
becomes an operator coupling total isospin $\mathcal{T} = \frac12$
and $\mathcal{T} = \frac32$ states as described in detail in 
\Ref~\cite{deltuva:05a}.
Instead of the breakup amplitude given by \Eq~\eqref{eq:UC1} 
we have to use the properly symmetrized form
\begin{subequations}
  \begin{gather}
  \begin{split} \label{eq:Uasyma}
    \langle \phi_0 (\mbf{p}_f \mbf{q}_f) & \nu_{0_f} |
    U_0 |\phi_\alpha (\mbf{q}_i) \nu_{\alpha_i} \rangle  \\ = {} &
    \sum_\sigma 
    \langle \phi_0 (\mbf{p}_f \mbf{q}_f) \nu_{0_f} |U_{0 \sigma}
    | \phi_\sigma (\mbf{q}_i) \nu_{\sigma_i} \rangle, 
  \end{split}
  \end{gather} \vspace{-5mm}
  \begin{gather}
  \begin{split}\label{eq:Uasymb}
    \langle \phi_0 & (\mbf{p}_f \mbf{q}_f)  \nu_{0_f} |
    U_0 |\phi_\alpha (\mbf{q}_i) \nu_{\alpha_i} \rangle  \\ = {} & 
    \lim_{R \to \infty} 
    \{ \zr(p_f) \langle \phi_0 (\mbf{p}_f \mbf{q}_f)   \nu_{0_f} | \\
      & \times   U_0^{(R)}(E_\alpha(q_i) + i0)   
       |\phi_\alpha (\mbf{q}_i) \nu_{\alpha_i} \rangle  \ZR(q_i) \}
  \end{split}
  \end{gather}
\end{subequations}
with $U_0^{(R)}(Z) = U^{(R)}_{0 \alpha}(Z) +
U^{(R)}_{0 \beta}(Z) P_{231} + U^{(R)}_{0 \gamma}(Z) P_{312}$
for the calculation of observables, 
$(\alpha \beta \gamma)$ being cyclic and $P_{231}$ and $P_{312}$ being the
two cyclic permutations of $(\alpha \beta \gamma)$.
The symmetrized breakup transition matrix
$U_0^{(R)}(Z)$ follows by quadrature
\begin{subequations} \label{eq:AGSsym}
\begin{gather} \label{eq:U0R}
  \begin{split}
    U_0^{(R)}(Z) = {} & (1+P) G_0^{-1}(Z) \\  & + 
    (1+P) T^{(R)}_{\alpha}(Z) G_0(Z) U^{(R)}(Z)
  \end{split}
\end{gather}
from the symmetrized multichannel transition matrix  
$U^{(R)}(Z) = U^{(R)}_{\alpha \alpha}(Z) +
U^{(R)}_{\alpha \beta}(Z) P_{231} + U^{(R)}_{\alpha \gamma}(Z) P_{312}$
of elastic $pd$ scattering, satisfying the standard
symmetrized form of the AGS integral equation \eqref{eq:Uba}, i.e.,
\begin{gather} \label{eq:UR}
  U^{(R)}(Z) = P G_0^{-1}(Z) + P T^{(R)}_{\alpha}(Z) G_0(Z) U^{(R)}(Z),
\end{gather}
\end{subequations}
with $P = P_{231} + P_{312}$.

The renormalization factors $\zR_{R}(q_i)$ and $z_R(p_f)$
in the initial and final channels are diverging phase factors 
defined in \Ref~\cite{taylor:74a} for a general screening
and calculated in \Refs~\cite{deltuva:05a,yamaguchi:03a}
for the screened Coulomb potential of \Eq~\eqref{eq:wr}, i.e.,
\begin{subequations} \label{eq:zrqp}
  \begin{align} \label{eq:zrq}
    \zR_{R}(q_i) = {} & e^{-2i \kappa(q_i)[\ln{(2q_i R)} - C/n]}, \\
    \label{eq:zrp}
    z_{R}(p_f) = {} & e^{-2i \kappa(p_f)[\ln{(2p_f R)} - C/n]},
  \end{align}
\end{subequations}
$\kappa(q_i) = \alpha_e M/q_i$ and $\kappa(p_f) = \alpha_e \mu /p_f$
being the $pd$ and $pp$ Coulomb  parameters, 
$M$ and $\mu$ the reduced $pd$ and $pp$ mass,
$C \approx 0.5772156649$ the Euler number,
and $n$ the exponent in \Eq~\eqref{eq:wr}.
In $pd$ elastic scattering, the renormalization factors were used in a
partial-wave dependent form  which yielded a slight advantage on
convergence with $R$ compared to the partial-wave independent
form \eqref{eq:zrqp}.
In breakup, the operator $T^{(R)}_{\alpha}(Z) G_0(Z) U^{(R)}(Z)$ in
\Eq~\eqref{eq:U0R} is calculated in a partial-wave basis, but
the  on-shell elements of the full breakup operator $U_0^{(R)}(Z)$
are calculated in a plane wave basis. Therefore 
the renormalization is only applicable in the partial-wave
independent form of \Eq~\eqref{eq:zrqp}.

The limit in \Eq~\eqref{eq:Uasymb} has to be performed numerically,
but, due to the finite-range nature of the breakup operator,
the infinite $R$ limit is reached with sufficient accuracy at 
rather modest screening radii $R$.
Furthermore, the longer-range pure Coulomb breakup part which after 
symmetrization reads $[1 + T_{\rho R}(Z) G_{0}(Z)] P v_{\alpha}
[1 + G^{(R)}_{\alpha}(Z) T^{\cm}_{\alpha R}(Z)]$
and the remaining shorter-range part can be renormalized with different
screening radii, since the limit in \Eq~\eqref{eq:Uasymb} exist for them
separately. 
The limit for the pure Coulomb breakup part can even be carried out
explicitly, since the renormalization of the screened Coulomb waves
yields the corresponding unscreened Coulomb waves accessible in 
configuration space; thus, the integral can be carried out numerically
in configuration space as was done indeed in \Ref~\cite{alt:94a}.
However, we find such a procedure unnecessary when our standard screening
function is used. In fact, in most cases there is even no necessity 
for splitting the full breakup amplitude into pure Coulomb and
Coulomb modified short-range parts, the only exception being 
the kinematical situations characterized by small
momentum transfer in the $pp$ subsystem which are sensitive to the Coulomb
interaction at larger distances.

The practical implementation of the outlined calculational scheme
faces a technical difficulty. We solve \Eq~\eqref{eq:UR}
in a partial-wave basis. The partial-wave expansion of the
screened Coulomb potential converges rather slowly.
In this context, the perturbation theory for higher two-baryon partial
waves developed in \Ref~\cite{deltuva:03b} 
is a very efficient and reliable technical tool for treating the screened 
Coulomb interaction in high partial waves. 
We vary the dividing line between partial waves included exactly and 
perturbatively  in order to test the convergence 
and thereby establish the validity of the procedure.
Furthermore, the partial-wave convergence becomes slightly faster
when replacing lowest-order screened Coulomb contributions in $U_0^{(R)}(Z)$
by the respective plane-wave results, i.e.,
\begin{gather}
U_0^{(R)}(Z) = [U_0^{(R)}(Z) - (1+P)w_{\alpha R} P] + (1+P)w_{\alpha R} P,
\end{gather}
where the first term converges with respect to partial waves faster
than $U_0^{(R)}(Z)$ itself and the second term is calculated 
\emph{without} partial-wave decomposition. 

With respect to the partial-wave expansion in the actual calculations of 
this paper, we obtain fully converged results by taking into account
the screened Coulomb interaction in two-baryon partial waves
with pair orbital angular momentum $L \le 15$;
orbital angular momenta $9 \le L \le 15$ can safely be treated 
perturbatively. The above values refer to the screening radius $R=30 \fm$;
for smaller screening radii the convergence in orbital angular momentum 
is faster. The hadronic interaction is taken into account
 in two-baryon partial waves with total angular momentum $I \le 5$. 
Both three-baryon total isospin $\mathcal{T} = \frac12$
and $\mathcal{T} = \frac32$ states are included.
The maximal three-baryon total angular momentum $\mathcal{J}$
considered is $\frac{61}{2}$.

\renewcommand{\scl}{0.62}
\begin{figure}[!]
\begin{center}
\includegraphics[scale=\scl]{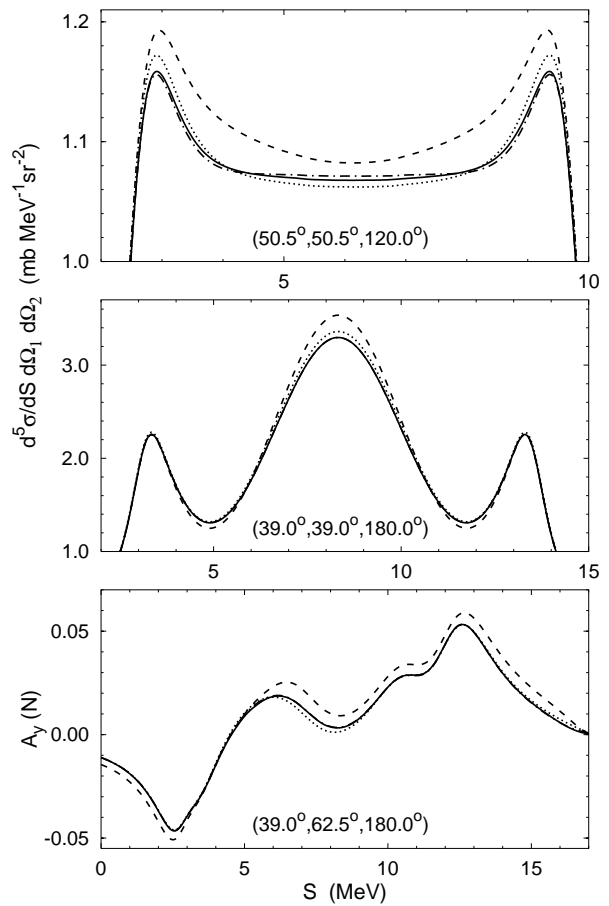}
\end{center}
\caption{\label{fig:R13}
Convergence of the $pd$ breakup  observables with screening radius $R$.
The differential cross section and the proton analyzing power $A_y(N)$ 
for $pd$ breakup at 13~MeV proton lab energy  are
shown as functions of the arclength $S$ along the kinematical curve.
Results for CD Bonn potential obtained with screening radius 
$R= 10$~fm (dotted curves), 20~fm (dash-dotted curves), and
30~fm (solid curves) are compared. Results without Coulomb (dashed curves)
are given as reference for the size of the Coulomb effect.}
\end{figure}
\renewcommand{\scl}{0.62}
\begin{figure}[!]
\begin{center}
\includegraphics[scale=\scl]{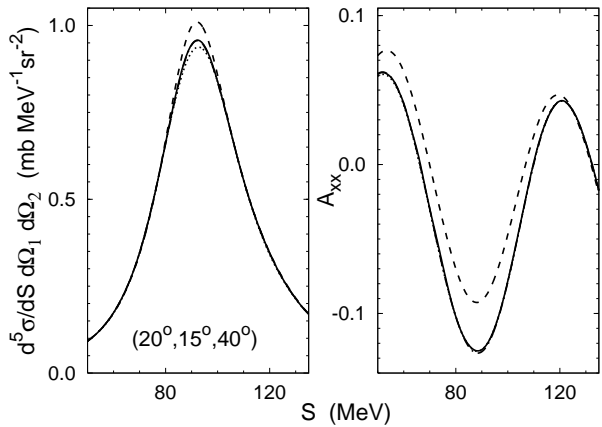}
\end{center}
\caption{\label{fig:R130d}
Convergence of the $pd$ breakup  observables with screening radius $R$.
The differential cross section and the deuteron 
analyzing power $A_{xx}$ for $pd$ breakup at 130~MeV are shown.
Curves as in \Fig~\ref{fig:R13}.}
\end{figure}
\begin{figure}[!]
\begin{center}
\includegraphics[scale=\scl]{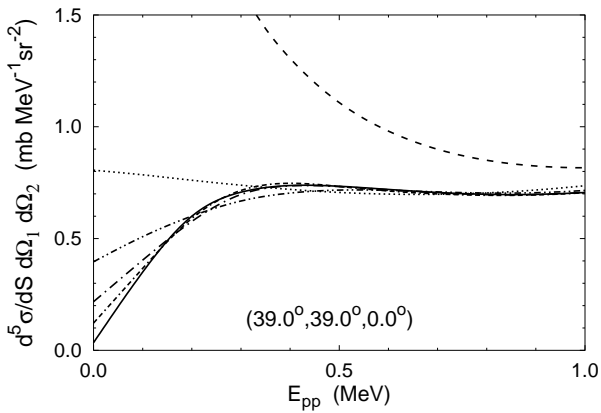}
\end{center}
\caption{\label{fig:R13ppfsi}
Convergence of the $pd$ breakup  observables with screening radius $R$.
The differential cross section  for $pd$ breakup at 13~MeV
proton lab energy in the $pp$-FSI configuration is shown
as function of the relative $pp$ energy $E_{pp}$.
Results obtained with screening radius $R= 10$~fm (dotted curves),
20~fm (dashed-double-dotted curves),
30~fm (dashed-dotted curves), 40~fm (double-dashed-dotted curves), and
60~fm (solid curves) are compared. Results without Coulomb (dashed curves)
are given as reference for the size of the Coulomb effect.}
\end{figure}
Figures~\ref{fig:R13} - \ref{fig:R13ppfsi} study the convergence
of our method with increasing screening
radius $R$ according to \Eq~\eqref{eq:Uasymb}.
All the calculations of this section are based on CD Bonn as the 
hadronic interaction.
The kinematical final-state configurations are characterized in a
standard way by the polar angles of the two protons and by the
azimuthal angle between them, $(\theta_1, \theta_2, \varphi_2 - \varphi_1)$.
We show several characteristic examples referring to $pd$ breakup
at 13 MeV proton lab energy and at 130 MeV deuteron lab energy.
The convergence is impressive for the spin-averaged differential cross 
section as well as for the spin observables in most kinematical 
situations as demonstrated in \Figs~\ref{fig:R13} and \ref{fig:R130d}.
The screening radius $R= 20 \fm$ is sufficient; only in the top plot
of \Fig~\ref{fig:R13} the curves for $R = 20 \fm$  and $R = 30 \fm$
 are graphically distinguishable.
The exception requiring larger screening radii 
is the differential cross section in kinematical
situations characterized by very low $pp$ relative energy $E_{pp}$, i.e.,
close to the $pp$ final-state interaction  ($pp$-FSI) regime,
as shown in \Fig~\ref{fig:R13ppfsi}. 
In there, the $pp$ repulsion is responsible for decreasing the cross section,
converting the $pp$-FSI peak obtained in the absence of Coulomb 
into a minimum with zero cross section at $p_f=0$, i.e., for $E_{pp}=0$.
A similar convergence problem also takes place
in $pp$ scattering at very low energies as discussed in 
\Ref~\cite{deltuva:05a}. In fact, screening and renormalization
procedure cannot be applied at $p_f=0$, since
the renormalization factor $z_R(p_f=0)$ is ill-defined.
Therefore an extrapolation has to be used to calculate
the observables at $p_f=0$, which works pretty well since
the observables vary smoothly with $p_f$. In \Fig~\ref{fig:R13ppfsi}
the fully converged result would start at zero for  $E_{pp}=0$.

The seen Coulomb effects and their physics implications are discussed
in \Sect~\ref{sec:res}.

\subsection{Three-body e.m. disintegration 
of $\He$ \label{sec:them}}

For the description of the considered e.m. processes the 
matrix element $\langle \psi^{(-)}_0 (\mbf{p}_f \mbf{q}_f) \nu_{0_f} |
  j^{\mu} (\mbf{Q}, \Kpl ) | B \rangle$ 
of the e.m. current operator
between the three-nucleon bound state and the breakup scattering state
has to be calculated.
The calculation of that matrix element without Coulomb and the meaning
of the momenta $\mbf{Q}$ and $\Kpl$ are discussed in
great length in \Refs~\cite{deltuva:04a,deltuva:04b}.
This subsection only discusses the modification which arises due to
the inclusion of the Coulomb interaction between the charged baryons.
Coulomb is included as a screened potential and the dependence of the
bound and scattering states, i.e., $| B^{(R)} \rangle$  and
$ |\psi^{(\pm)(R)}_0 (\mbf{p}_f \mbf{q}_f) \nu_{0_f} \rangle $,
on the screening radius $R$ is notationally made explicit.
In analogy to $pd$ breakup,
the current matrix element referring to the unscreened Coulomb potential
is obtained via renormalization of the matrix element 
referring to the screened Coulomb potential in the infinite $R$ limit
\begin{gather} \label{eq:jR}
  \begin{split}
    \langle \psi^{(-)}_0 & (\mbf{p}_f \mbf{q}_f)  \nu_{0_f} |
    j^{\mu} (\mbf{Q}, \Kpl ) | B \rangle \\ = {} & 
    \lim_{R \to \infty} 
    \{ \zr(p_f) \langle \psi^{(-)(R)}_0 (\mbf{p}_f \mbf{q}_f) \nu_{0_f} |
    j^{\mu} (\mbf{Q}, \Kpl ) | B^{(R)} \rangle  \}.
  \end{split}
\end{gather}
The renormalization factor $z_R(p_f)$ is the same as used in $pd$ breakup
for the final state. Due to the short-range nature of 
$ j^{\mu} (\mbf{Q}, \Kpl ) | B^{(R)} \rangle$ 
the limit $R \to \infty$ is reached with sufficient accuracy at 
finite screening radii $R$. 
The presence of the bound-state wave function in the matrix element
strongly suppresses the contribution of the screened Coulomb interaction 
in high partial waves, i.e., two-baryon partial waves with orbital 
angular momentum $L \le 6$ are sufficient for convergence.
The other quantum-number related cutoffs in the partial-wave dependence
of the matrix element are the same as in \Refs~\cite{deltuva:04a,deltuva:04b},
i.e.,  $I \le 4$, $\mathcal{J} \le \frac{15}{2}$ for photoreactions, 
and  $I \le 3$, $\mathcal{J} \le \frac{35}{2}$ for inelastic
electron scattering from $\He$. All calculations include
both total isospin $\mathcal{T} = \frac12$
and $\mathcal{T} = \frac32$ states.

\begin{figure}[!]
\begin{center}
\includegraphics[scale=\scl]{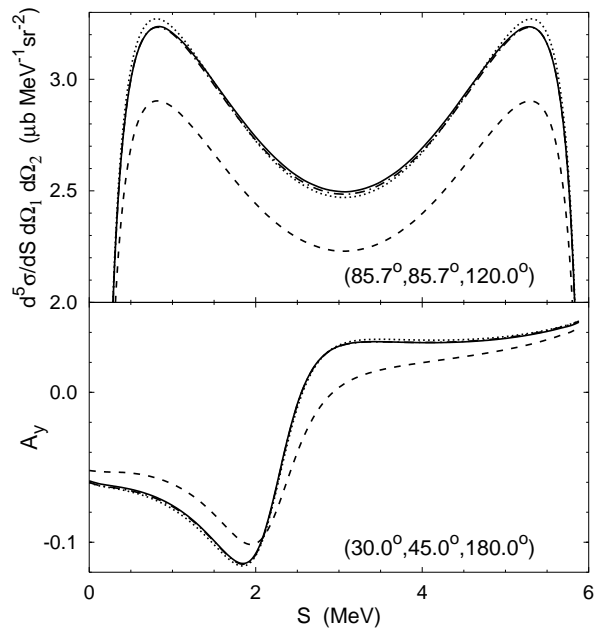}
\end{center}
\caption{\label{fig:Rg15}
Convergence of the $\He(\gamma,pp)n$ reaction observables 
with screening radius $R$.
The differential cross section and the target analyzing power $A_y$
at 15~MeV photon lab energy are shown.
Curves as in \Fig~\ref{fig:R13}.}
\end{figure}

\begin{figure}[!]
\begin{center}
\includegraphics[scale=\scl]{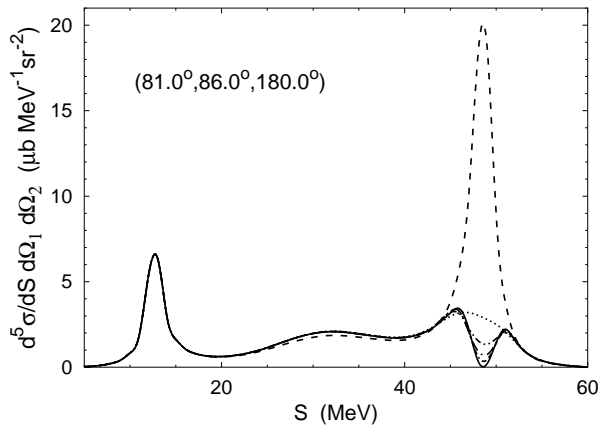}
\end{center}
\caption{\label{fig:Rg55}
Convergence of the $\He(\gamma,pn)p$ reaction observables 
with screening radius $R$.
The differential cross section at 55~MeV photon lab energy in the 
$pp$-FSI configuration is shown.
Curves as in \Fig~\ref{fig:R13ppfsi}.}
\end{figure}

Figures \ref{fig:Rg15} and  \ref{fig:Rg55} study the convergence
of our method with increasing screening radius $R$ for the 
three-body photodisintegration of $\He$ at 15 MeV and 55 MeV
photon lab energy.
The calculations are again based on CD Bonn as the hadronic interaction
and the currents from \Refs~\cite{deltuva:04a,deltuva:04b}.
We show the differential cross section and the target analyzing power $A_y$
for selected kinematic configurations. 
The convergence is again extremely good and quite comparable to $pd$ breakup;
the screening radius $R= 20 \fm$ is fully sufficient in most cases.
The only exceptional cases, as in the $pd$ breakup, are 
$pp$-FSI regimes as shown in \Fig~\ref{fig:Rg55}.
The convergence with increasing screening radius $R$ is the same for
three-body electrodisintegration of $\He$; we therefore omit a
corresponding figure.

\section{Results \label{sec:res}}

We base our calculations on the two-baryon coupled-channel potential
CD Bonn + $\Delta$ with and without Coulomb and use the CD Bonn potential
with  Coulomb as purely nucleonic reference. We use the charge
and current operators of \Refs~\cite{deltuva:04a,deltuva:04b},
appropriate for the underlying dynamics. In contrast to  
\Ref~\cite{deltuva:05a} we do not include one-nucleon relativistic 
charge corrections for photoreactions, since their effect on the considered
observables is very small.

Obviously, we have much more predictions than it is possible to show. 
Therefore we make a selection of the most interesting predictions 
which illustrate the message we believe the results tell us.
The readers are welcome to obtain the results
for their favorite data from us.

\subsection{Proton-deuteron breakup \label{sec:pdb}}

Figures \ref{fig:d5sss} - \ref{fig:d5sfsi} give our results for the
fivefold differential cross section at 10.5 MeV, 13 MeV, 19 MeV, and 
65 MeV proton lab energies in the standard space star, collinear,
quasifree scattering (QFS), and $np$ final-state interaction ($np$-FSI)
configurations, for which there is available experimental data.
Though the inclusion of Coulomb slightly improves the agreement with
data in the space star configurations in \Fig~\ref{fig:d5sss},
the Coulomb effect is far too small
to reproduce the difference between $pd$ and $nd$ data and
to resolve the so-called \emph{space star anomaly} at 13 MeV.
The inclusion of Coulomb clearly improves the description of the data 
around the collinear points at lower energies, i.e., at the minima 
in  \Fig~\ref{fig:d5scl}.
The remaining discrepancies around the peaks are probably due to the
finite geometry, not taken into account in our calculations
owing to the lack of information on experimental details,
but may also be due to the underlying hadronic interaction.
The inclusion of Coulomb decreases the differential cross section
around the QFS peaks, i.e., around the central peaks in \Fig~\ref{fig:d5sqfs}; 
those changes are supported by the data at lower energies. 
In the $np$-FSI configurations of \Fig~\ref{fig:d5sfsi}
the Coulomb effect is rather insignificant.

\renewcommand{\scl}{0.62}
\begin{figure}[!]
\begin{center}
\includegraphics[scale=\scl]{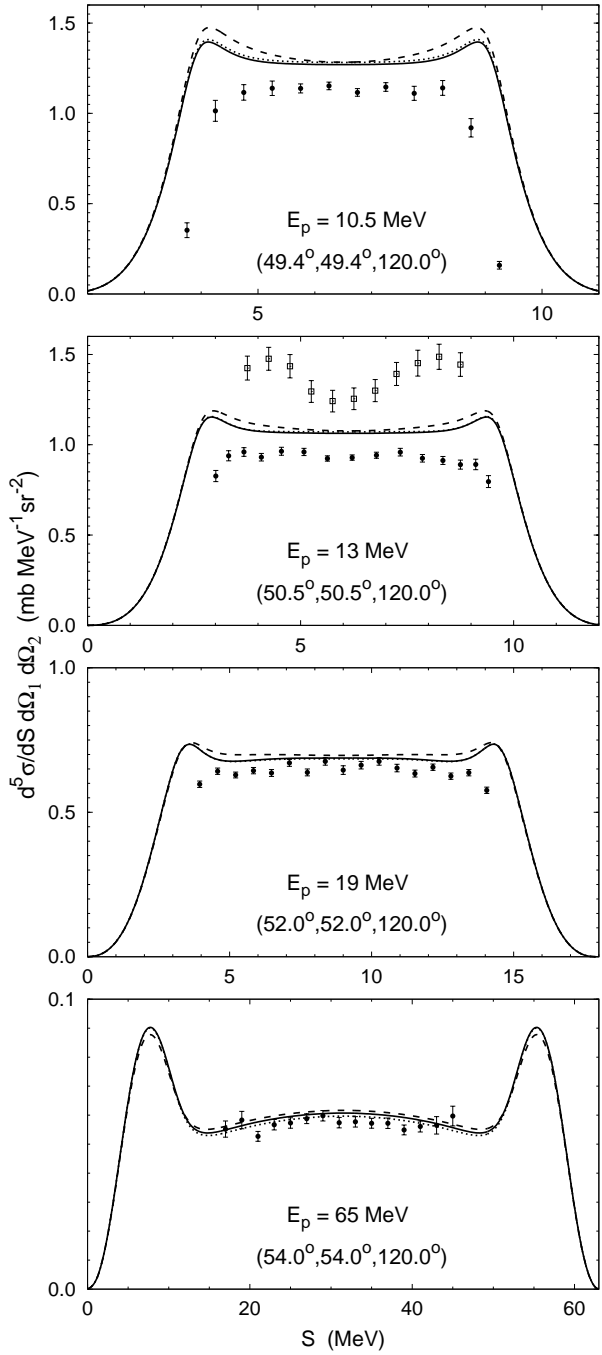}
\end{center}
\caption{\label{fig:d5sss}
Differential cross section for space star configurations 
 as function of the arclength $S$ along the kinematical curve.
Results including $\Delta$-isobar excitation and the Coulomb interaction
(solid curves) are compared to results without Coulomb (dashed curves).
In order to appreciate the size of the $\Delta$-isobar effect, the purely
nucleonic results including Coulomb are also shown (dotted curves).
The experimental $pd$ data (circles) are from \Ref~\cite{grossmann:96} 
at 10.5 MeV, from \Ref~\cite{rauprich:91} at 13 MeV, 
from \Ref~\cite{patberg:96} at 19 MeV, from \Ref~\cite{zejma:97a} at 65 MeV,
and $nd$ data at 13 MeV (squares) from \Ref~\cite{strate:89}.}
\end{figure}

\begin{figure}[!]
\begin{center}
\includegraphics[scale=\scl]{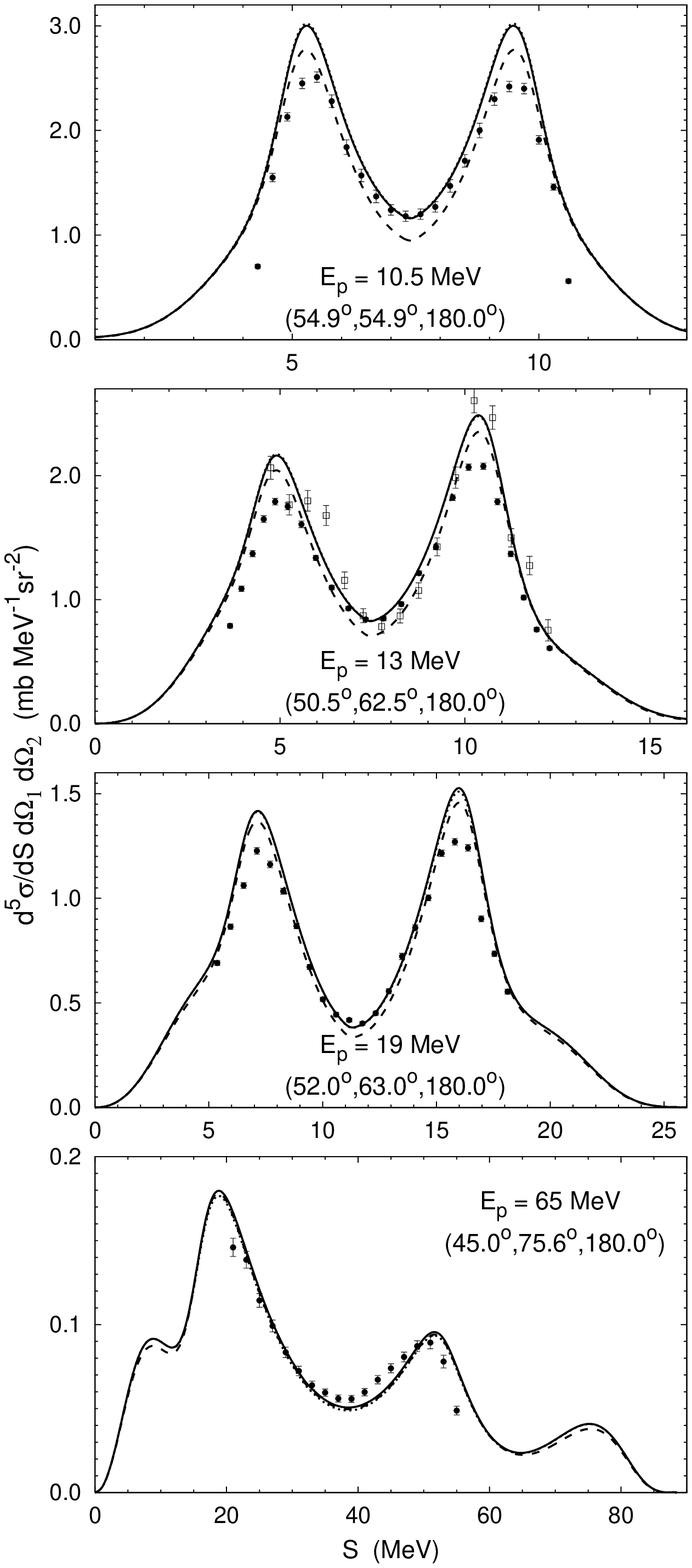}
\end{center}
\caption{\label{fig:d5scl}
Differential cross section for collinear configurations.
Curves and experimental data as in \Fig~\ref{fig:d5sss}, except for
65 MeV data from \Ref~\cite{allet:94a}.}
\end{figure}

\begin{figure}[!]
\begin{center}
\includegraphics[scale=\scl]{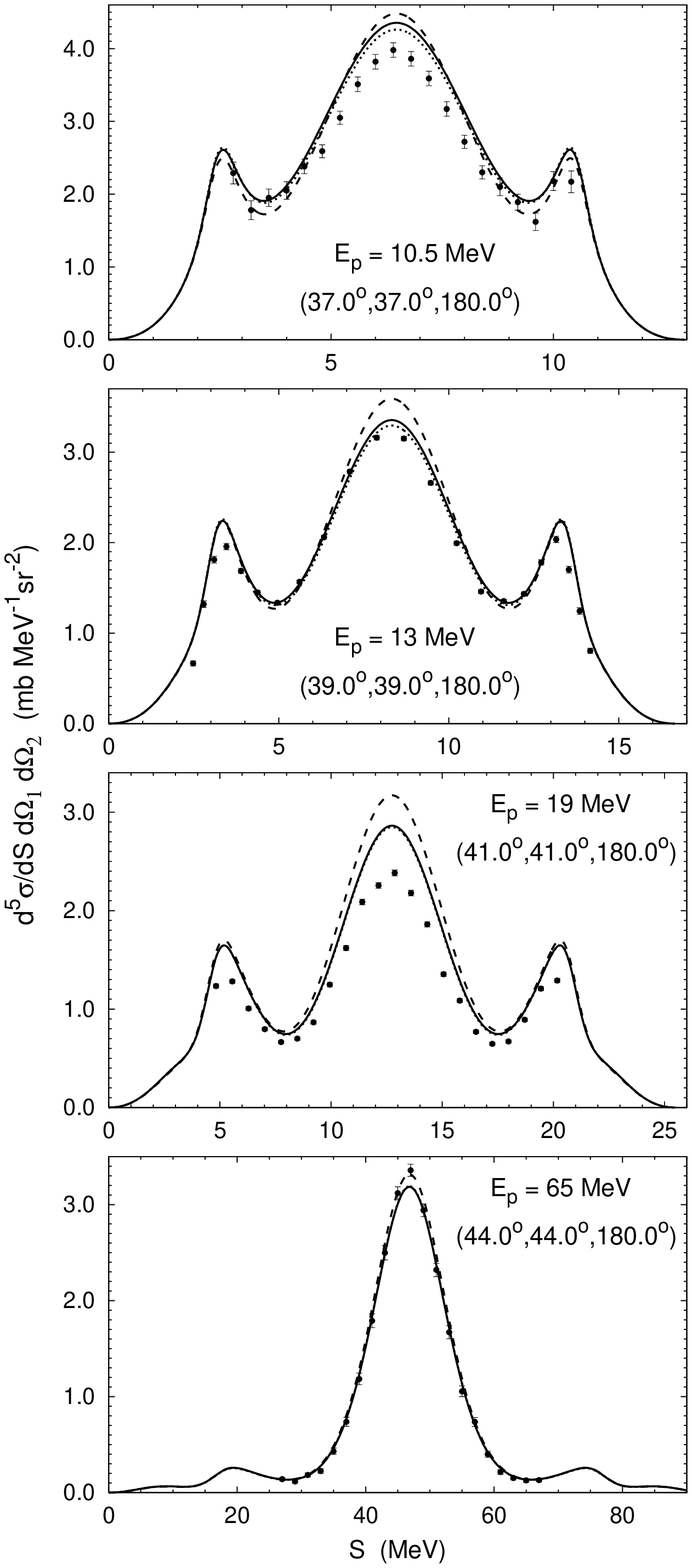}
\end{center}
\caption{\label{fig:d5sqfs}
Differential cross section for QFS configurations.
Curves and experimental data as in \Fig~\ref{fig:d5sss}, except for
65 MeV data from \Ref~\cite{allet:96a}.}
\end{figure}

\begin{figure}[!]
\begin{center}
\includegraphics[scale=\scl]{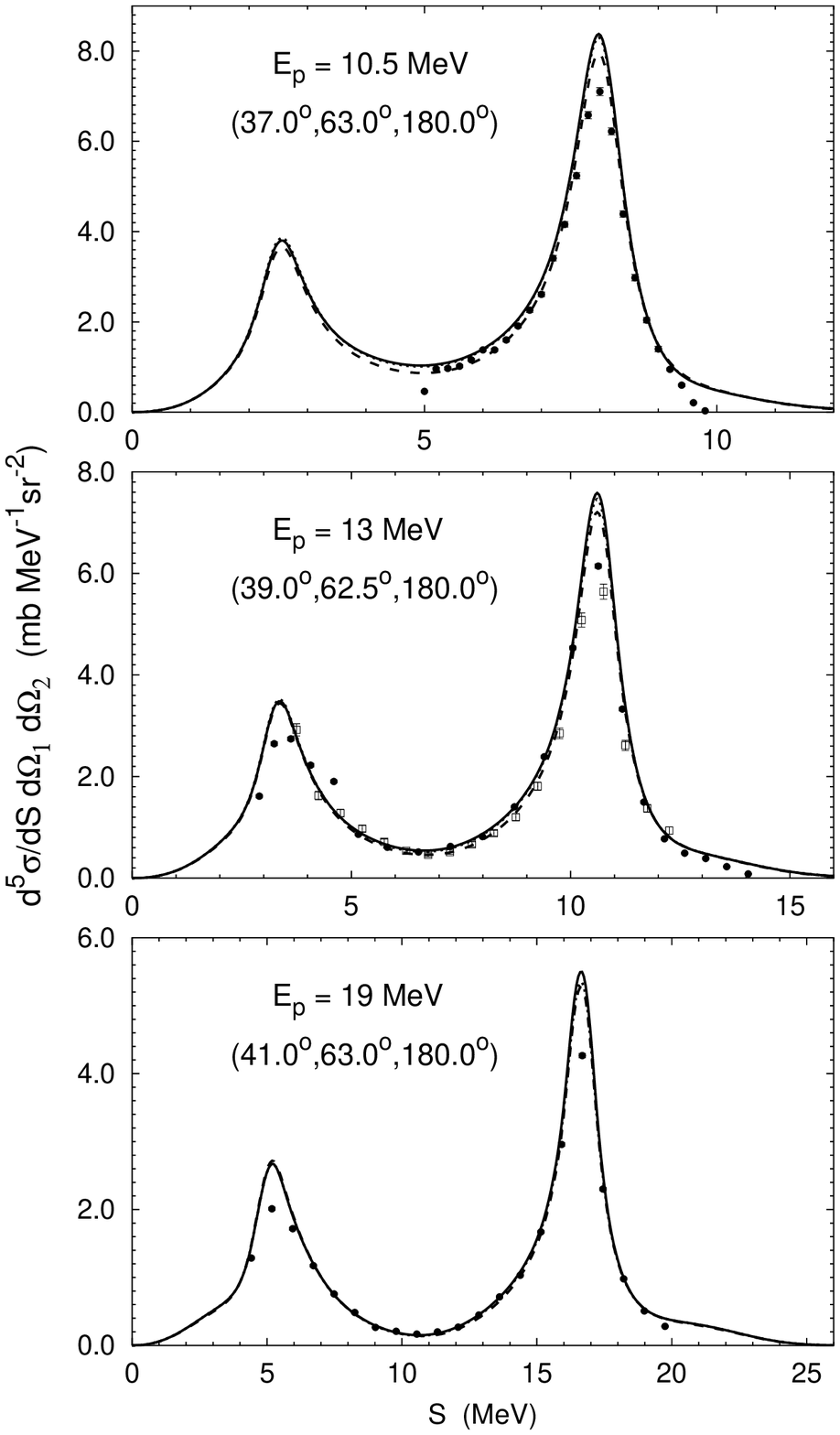}
\end{center}
\caption{\label{fig:d5sfsi}
Differential cross section for $np$-FSI configurations.
Curves and experimental data as in \Fig~\ref{fig:d5sss}.}
\end{figure}

\begin{figure}[!]
\begin{center}
\includegraphics[scale=\scl]{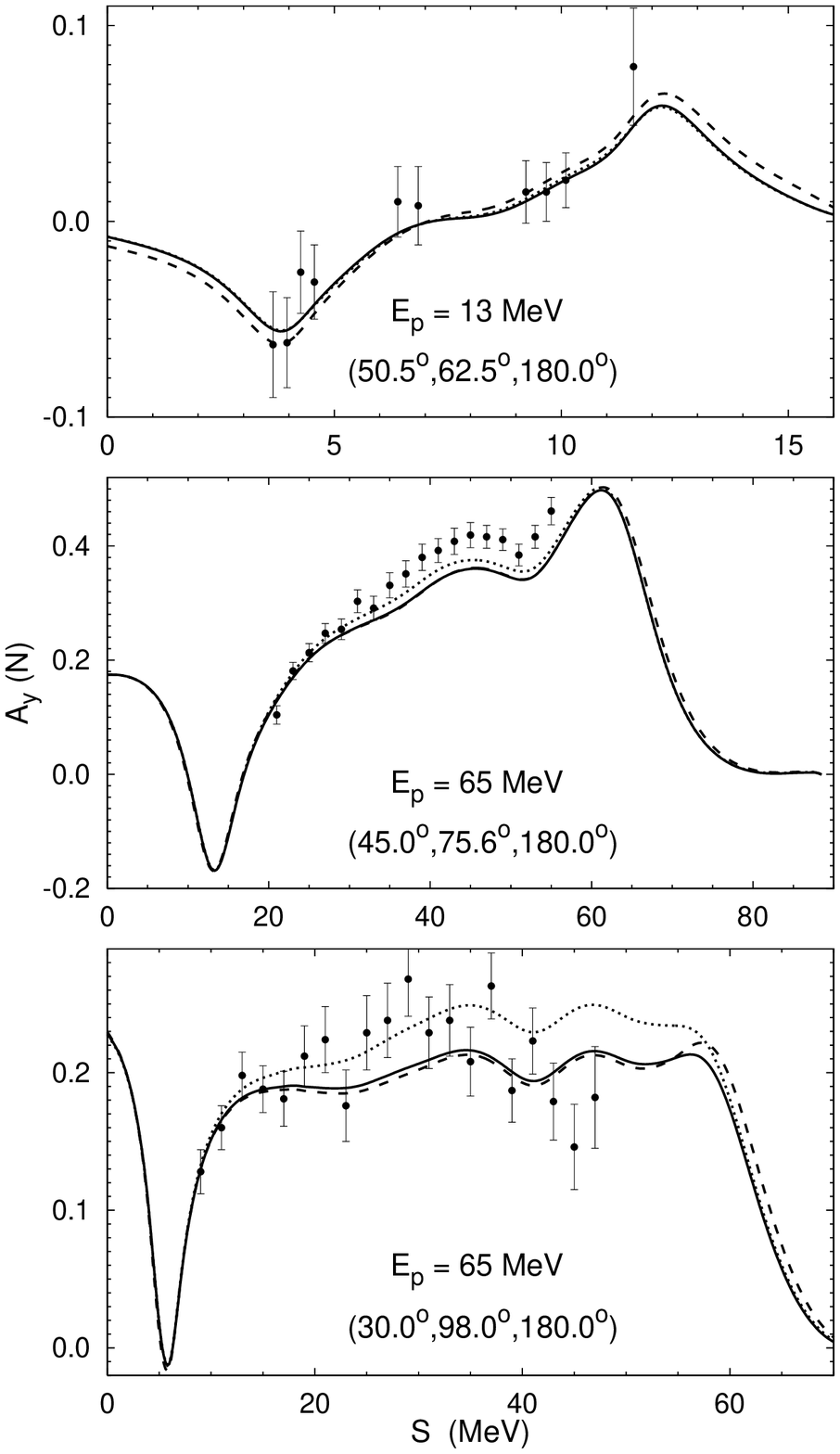}
\end{center}
\caption{\label{fig:Aycl}
Proton analyzing  power for collinear configurations at 13 MeV and 65 MeV
proton lab energy.
Curves and experimental data as in \Fig~\ref{fig:d5scl}.}
\end{figure}

The Coulomb effect on proton analyzing powers in the considered 
kinematical configurations is usually small on
the scale of the experimental error bars. 
We therefore show in  \Fig~\ref{fig:Aycl} only few collinear configurations.

\begin{figure}[!]
\begin{center}
\includegraphics[scale=\scl]{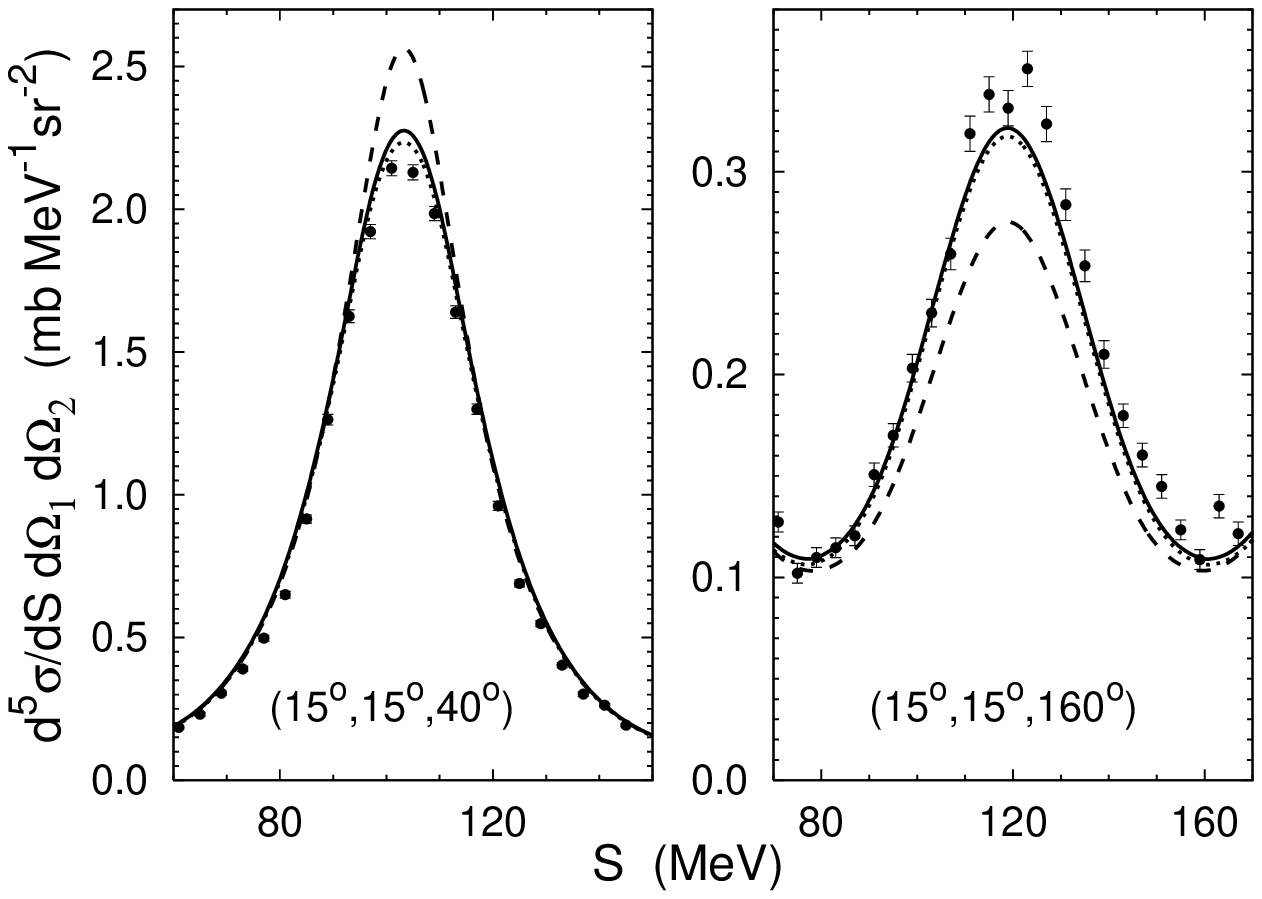}
\end{center}
\caption{\label{fig:d5s130d}
Differential cross section for $pd$ breakup at 130 MeV deuteron lab energy.
Curves as in \Fig~\ref{fig:d5sss}.
The experimental data are from \Ref~\cite{kistryn:05a}. }
\end{figure}

\begin{figure}[!]
\begin{center}
\includegraphics[scale=\scl]{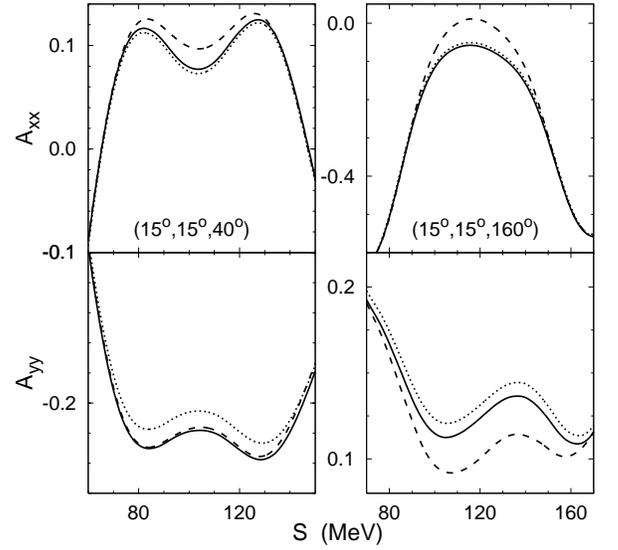}
\end{center}
\caption{\label{fig:A130d}
Deuteron analyzing powers for $pd$ breakup at 130 MeV deuteron lab energy.
Curves as in \Fig~\ref{fig:d5sss}. }
\end{figure}

\begin{figure}[!]
\begin{center}
\includegraphics[scale=\scl]{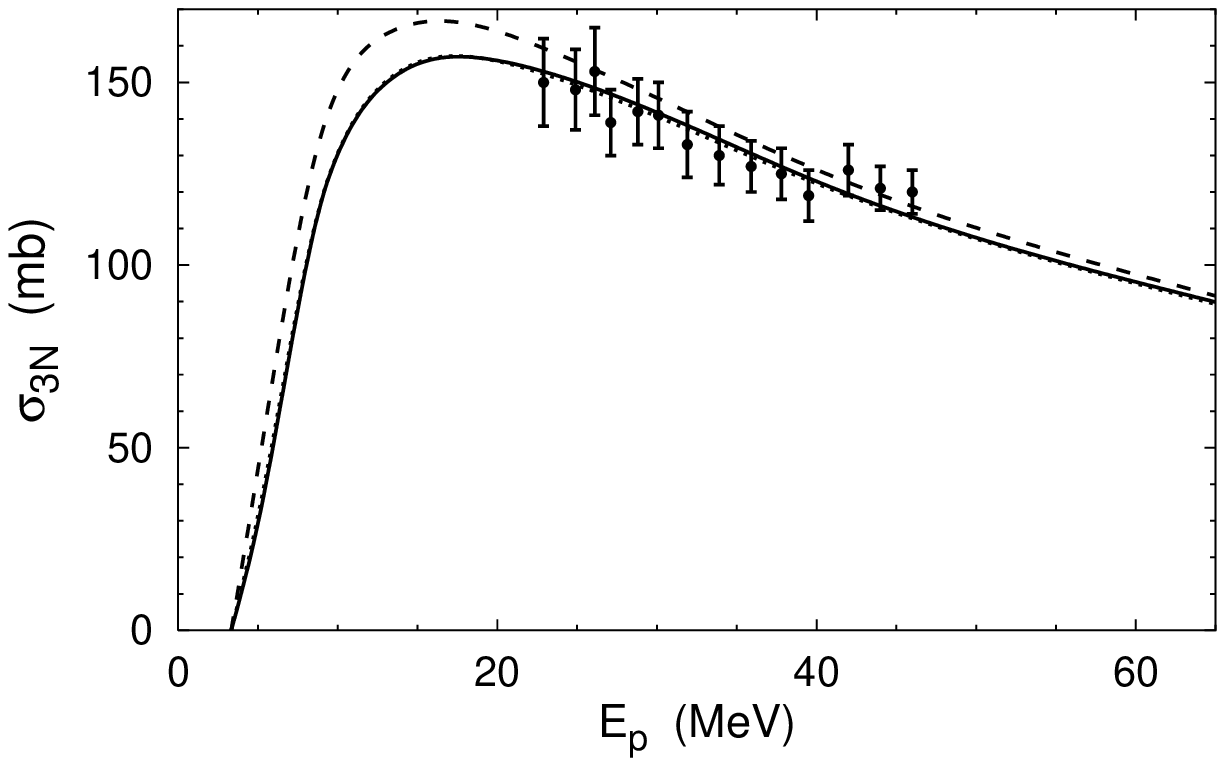}
\end{center}
\caption{\label{fig:tot3N}
Total cross section for $pd$ breakup as function of the proton lab energy.
Curves as in \Fig~\ref{fig:d5sss}.
The experimental data are from \Ref~\cite{carlson:73a}. }
\end{figure}

Recently,  $pd$ breakup has been measured at 130 MeV deuteron lab energy 
in a  variety of kinematical configurations \cite{kistryn:05a}.
In some of them we find significant Coulomb effects for the differential
cross section as well as for the deuteron analyzing powers. Examples are shown
in \Figs~\ref{fig:d5s130d} and \ref{fig:A130d}.
By and large the agreement between theoretical predictions and 
experimental data is improved. 
The $pp$-FSI repulsion is  responsible for lowering the peak of the 
differential cross section in the configuration 
$(15^{\circ},15^{\circ},40^{\circ})$ in \Fig~\ref{fig:d5s130d} left, 
where the relative $pp$ energy is rather low at the peak. In contrast,
the relative $pp$ energy gets considerably increased as one 
changes the azimuthal angle to $160^{\circ}$ in \Fig~\ref{fig:d5s130d} right,
leading to an increase of the differential cross section due to Coulomb.
Since the total breakup cross section at this energy,
corresponding to 65 MeV proton lab energy, is almost unaffected 
by Coulomb, as shown in \Fig~\ref{fig:tot3N}, one may expect in given 
configurations an increase of the cross section due to Coulomb
to compensate for the sharp decrease of the cross section in the vicinity
of $pp$-FSI points.
Figure \ref{fig:A130d} shows deuteron tensor analyzing powers $A_{xx}$
and $A_{yy}$ with moderate Coulomb effect for the same configurations
for which experimental data may become available soon \cite{stephan:05a}.

Compared to the results of \Ref~\cite{alt:94a} based on a simple hadronic
$S$-wave potential, we see a rough qualitative agreement in most cases.
Quantitatively, the Coulomb effect we observe is smaller than the one
of \Ref~\cite{alt:94a}.

Figures \ref{fig:d5sss} - \ref{fig:tot3N} recall also the 
$\Delta$-isobar effect on observables, which, in most cases we studied,
is much smaller than the Coulomb effect.
As expected, the $\Delta$-isobar effect on polarization observables is
more significant than on the differential cross sections, which confirms
previous findings \cite{deltuva:03c}.

\subsection{Three-body e.m. disintegration of $\He$}

\begin{figure}[!]
\begin{center}
\includegraphics[scale=\scl]{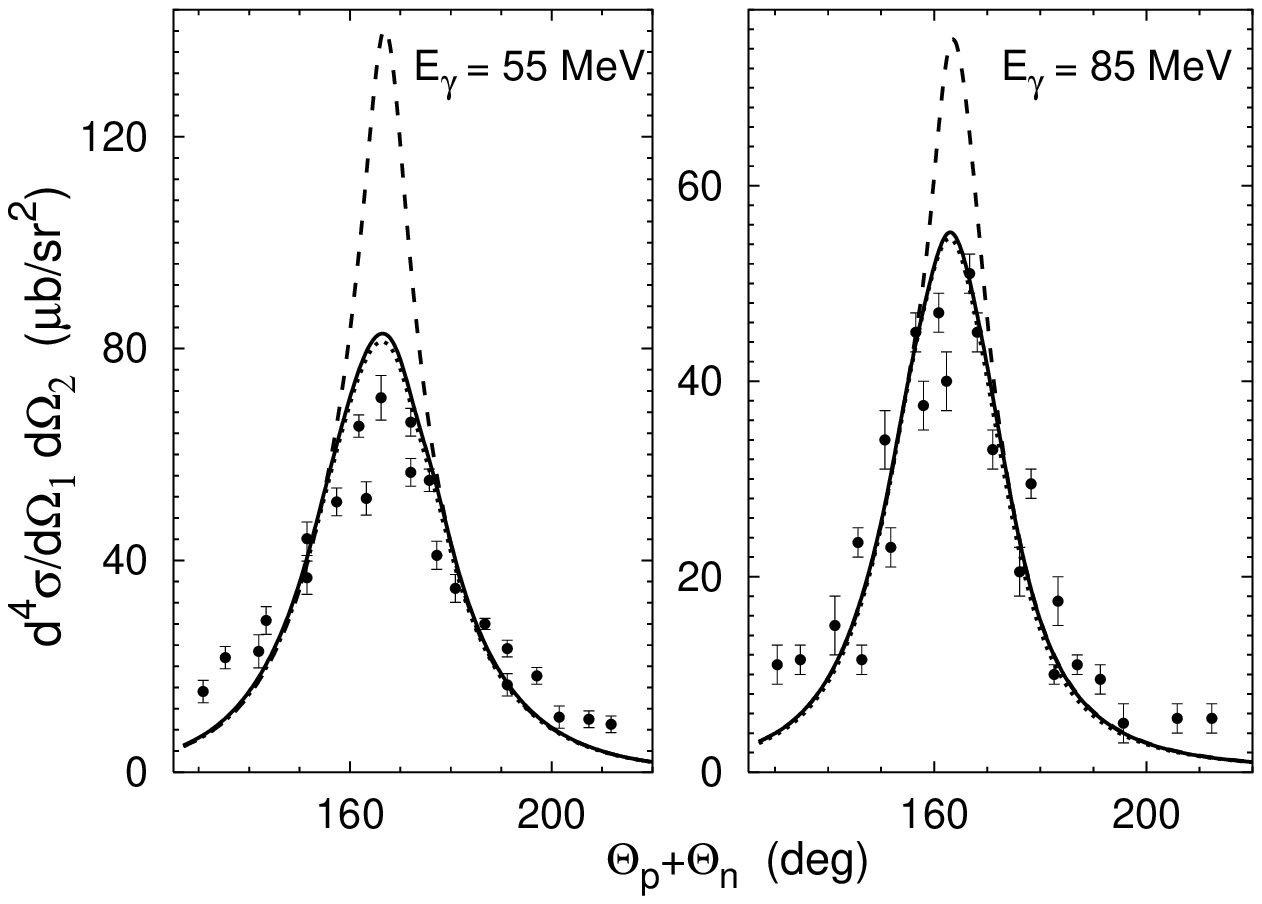}
\end{center}
\caption{\label{fig:d4s}
The  semiinclusive fourfold
differential cross section for $\He(\gamma,pn)p$ reaction
 at 55 MeV and 85 MeV photon lab energy as function of the $np$ opening
angle $\theta_p + \theta_n$ with $\theta_p = 81^{\circ}$.
Curves as in \Fig~\ref{fig:d5sss}.
The experimental data are from \Ref~\cite{kolb:91a}. }
\end{figure}

\begin{figure}[!]
\begin{center}
\includegraphics[scale=\scl]{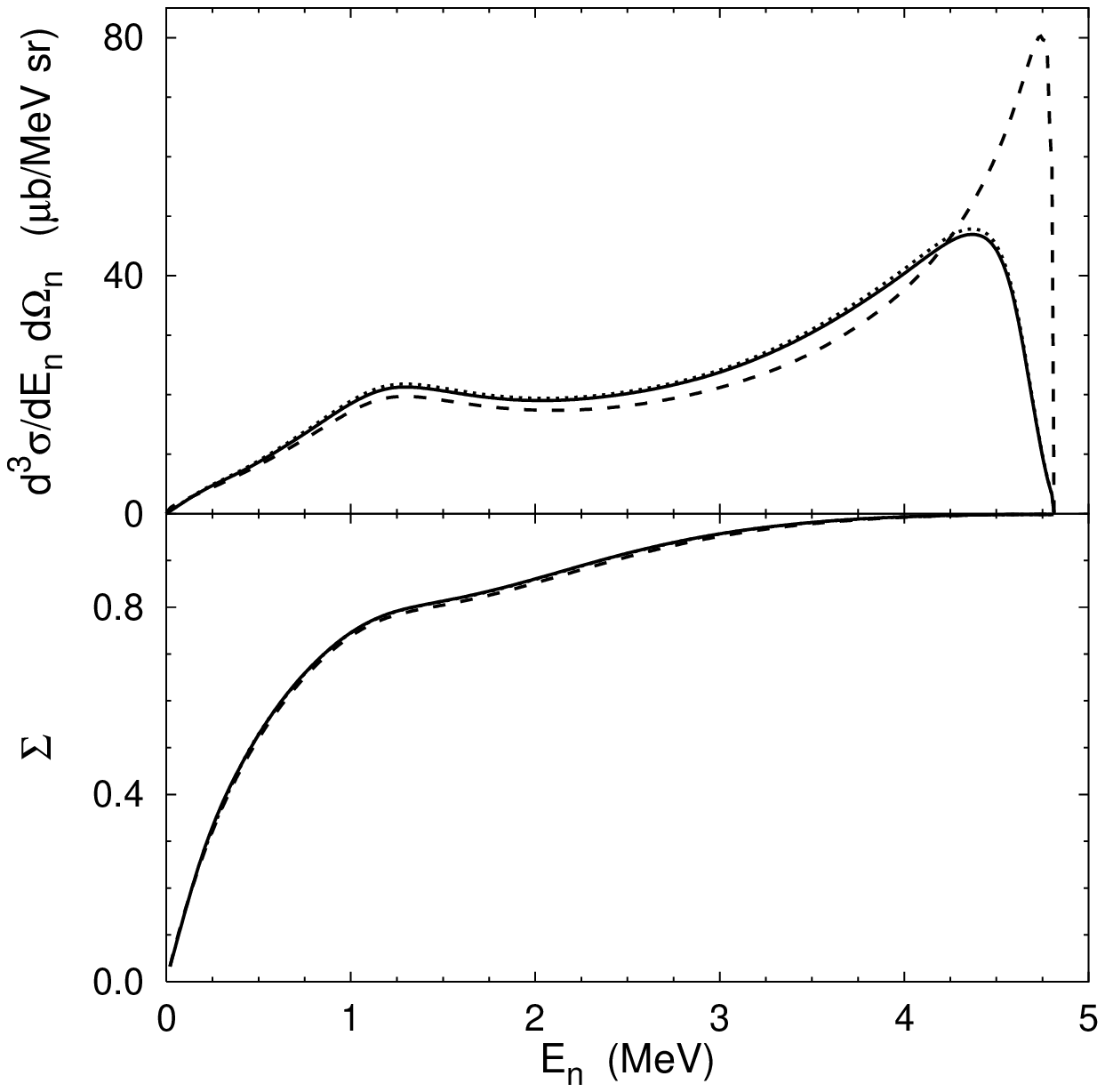}
\end{center}
\caption{\label{fig:d3s}
The  semiinclusive threefold
differential cross section and photon analyzing power $\Sigma$
for $\He(\vec{\gamma},n)pp$ reaction
 at 15 MeV photon lab energy as function of the neutron energy
$E_n$ for the neutron scattering angle $\theta_n = 90^{\circ}$.
Curves as in \Fig~\ref{fig:d5sss}.}
\end{figure}

Experimental data for three-body photodisintegration of $\He$ are very scarce;
we therefore show in \Fig~\ref{fig:d4s} only two examples referring to
the semiinclusive $\He(\gamma,pn)p$ reaction at 55 MeV and 85 MeV photon 
lab energy.
The semiinclusive fourfold differential cross section is obtained
from the standard fivefold differential cross section by integrating over
the kinematical curve $S$. For scattering angles corresponding to the
peak of the fourfold differential cross section the region of the 
phase space to be integrated over contains $pp$-FSI regime
where the $pp$-FSI peak obtained without Coulomb is converted into a
minimum as shown in \Fig~\ref{fig:Rg55}. Therefore the fourfold differential 
cross section in \Fig~\ref{fig:d4s}
is also significantly reduced by the inclusion of Coulomb,
clearly improving the agreement with the data.
A similar Coulomb effect of the same origin is shown in \Fig~\ref{fig:d3s}
for the  semiinclusive threefold differential cross section for
$\He(\vec{\gamma},n)pp$ reaction  at 15 MeV photon lab energy.
In contrast, the photon analyzing power remains almost unchanged by the
inclusion of Coulomb. The experiment measuring this reaction is in 
progress~\cite{tornow:05a}, but the data are not available yet.
Again the importance of $\Delta$-isobar degree of freedom is considerably
smaller than the effect of Coulomb.

\begin{figure}[!]
\begin{center}
\includegraphics[scale=\scl]{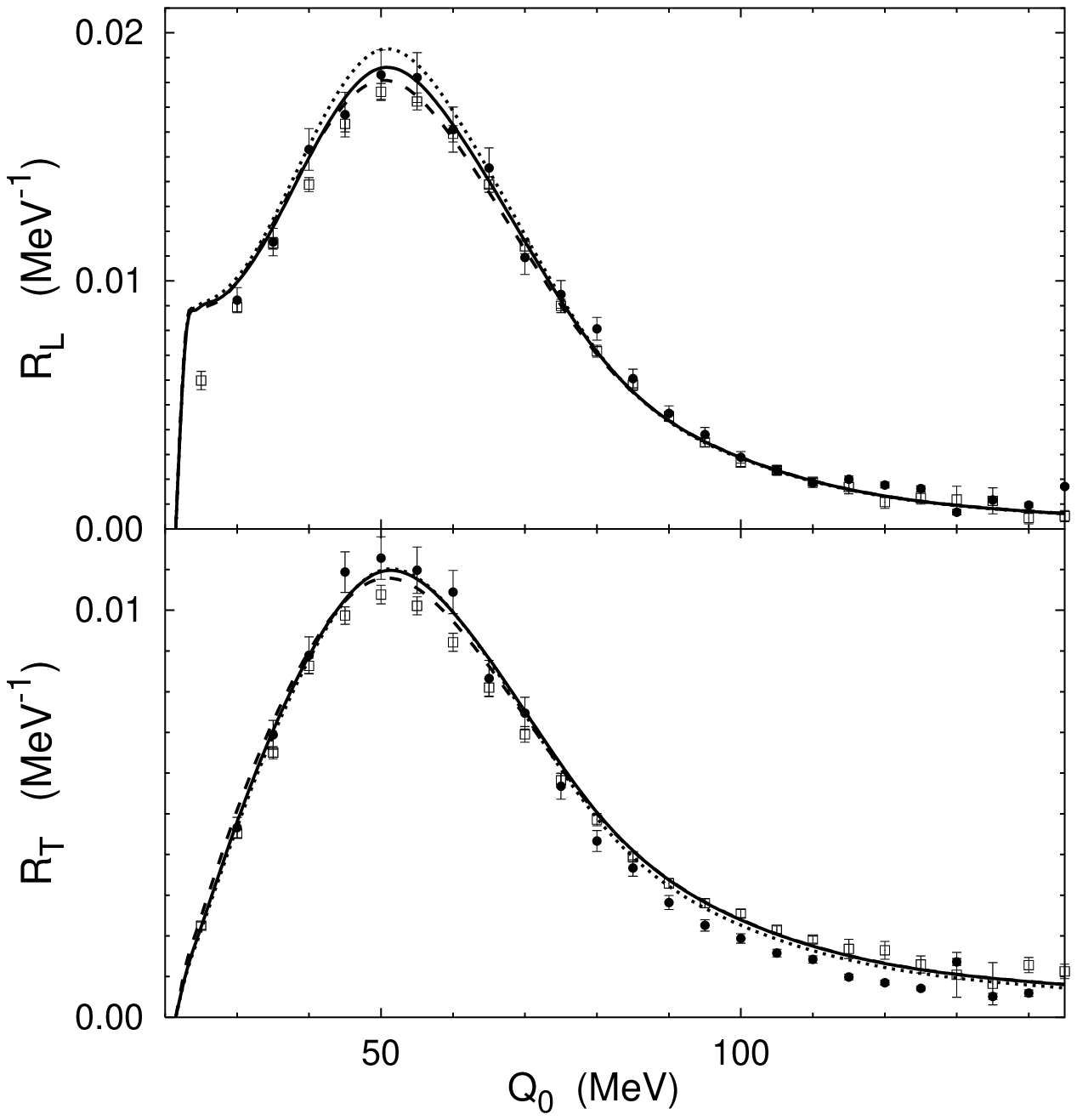}
\end{center}
\caption{\label{fig:RF300}
$\He$ inclusive longitudinal and transverse response
functions $R_L$ and $R_T$
for the momentum transfer $\;|\mbf{Q}| = 300\;\mathrm{MeV}$ as functions
of the energy transfer $Q_0$.
Curves as in \Fig~\ref{fig:d5sss}.
The experimental data are from \Ref~\cite{dow:88a} (circles)
and from \Ref~\cite{marchand:85a} (squares).}
\end{figure}

\begin{figure}[!]
\begin{center}
\includegraphics[scale=\scl]{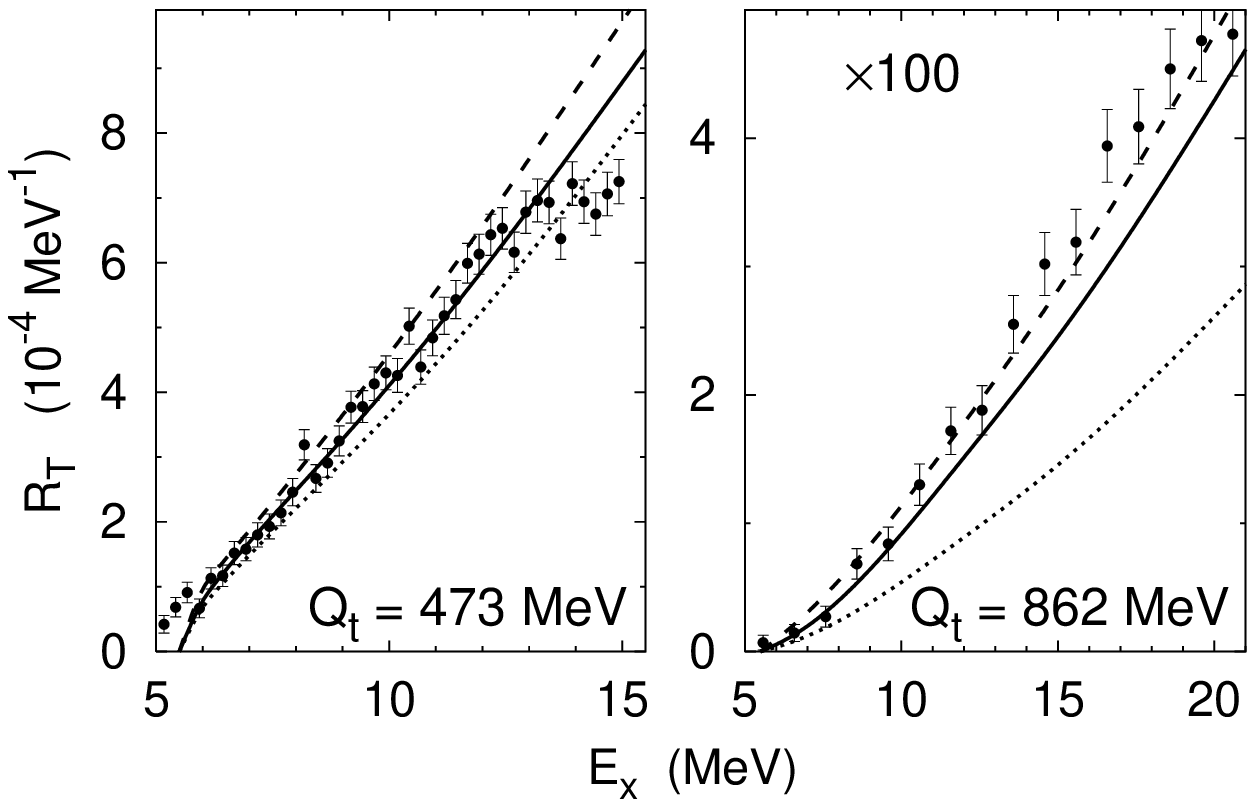}
\end{center}
\caption{\label{fig:RTt}
$\He$ inclusive transverse response
function $R_T$ near threshold as function of the excitation energy $E_x$,
$Q_t$ being the value of three-momentum transfer at threshold.
Curves as in \Fig~\ref{fig:d5sss}.
The experimental data are from \Ref~\cite{hicks:03a}.}
\end{figure}

The available data of three-nucleon electrodisintegration of $\He$ refer
to fully inclusive observables. In \Fig~\ref{fig:RF300} we show
$\He$ inclusive longitudinal and transverse response
functions $R_L$ and $R_T$ as examples. Though the Coulomb effect may be
large in particular kinematic regions, it is rather insignificant
for the total cross section and therefore also for response functions.
Only the transverse response function near threshold is affected
more visibly as shown in \Fig~\ref{fig:RTt}; at higher momentum transfer 
there is also quite a large $\Delta$-isobar effect.

\section{Summary \label{sec:concl}}

In this paper we show how the Coulomb interaction between the charged baryons
can be included into the momentum-space description of
proton-deuteron breakup and of three-body e.m. disintegration of $\He$
using the screening and renormalization approach.
The theoretical framework is the AGS integral equation \cite{alt:67a}.
The calculations are done on the same level of accuracy and
sophistication as for the corresponding neutron-deuteron and $\Hh$ reactions.
The conclusions of the paper refer to the developed technique and
to the physics results obtained with that technique.

\emph{Technically}, the idea of screening and renormalization is the one of
\Refs~\cite{taylor:74a,alt:78a,alt:94a}.
However, our practical realization  differs quite significantly
from the one of \Ref~\cite{alt:94a}:

(1) We use modern hadronic interactions, CD Bonn and CD Bonn + $\Delta$,
in contrast to the simple $S$-wave separable potentials
of \Ref~\cite{alt:94a}. Our use of the full  potential
requires the standard form of the three-particle equations,
different from the quasiparticle approach of \Ref~\cite{alt:94a}.

(2) We do not approximate the screened Coulomb transition
matrix by the screened Coulomb potential.

(3) The quasiparticle approach of \Ref~\cite{alt:94a} treats the
screened Coulomb potential between the protons without partial-wave
expansion and therefore has no problems with the slow convergence
of that expansion. Our solution of three-nucleon equations proceeds
in partial-wave basis and therefore faces the slow partial-wave
convergence of the Coulomb interaction between the charged baryons.
However, we are able to
obtain fully converged results by choosing a special form of the screening
function and by using the perturbation theory of \Ref~\cite{deltuva:03b}
for treating the screened Coulomb transition matrix in high partial waves.
This would not be possible, if we had used Yukawa screening as in
\Ref~\cite{alt:94a} for two reasons: (a) The convergence
with respect to screening  would require much larger radii $R$;
(b) The larger values of $R$ would necessitate the solution of the
AGS equation with much higher angular momentum states.

(4) Our method for including the Coulomb interaction is efficient.
Though the number of the isospin triplet partial waves to be taken into account
is considerably higher than in the case without Coulomb,
the required computing time increases only by a factor of 3 --- 4
for each screening radius $R$,
due to the use of perturbation theory for high partial waves.

The obtained results are fully converged with respect to the
screening and with respect to the quantum number cutoffs; they are
therefore well checked for their validity.
The employed technique gets cumbersome in kinematical regions
with very low relative $pp$ energy,
i.e., $pp$ c.m. energies below 0.1~MeV, due to the need for quite large
screening radii. 

\emph{Physicswise}, the Coulomb effect in $pd$ breakup
and in three-body e.m. disintegration of $\He$
is extremely important in kinematical regimes close to $pp$-FSI.
There the $pp$ repulsion converts the $pp$-FSI peak obtained in the 
absence of Coulomb into a minimum with zero cross section.
This significant change of the cross section behavior has important
consequences in nearby configurations where one may observe
instead an increase of the cross section due to Coulomb.
This phenomenon is independent of the beam energy and depends solely
on specific momentum distributions of the three-nucleon final state.
Therefore, unlike in $pd$ elastic scattering where the Coulomb
contribution decreases with the beam energy until it gets confined
to the forward direction, in three-body breakup large Coulomb effects
may always be found in specific configurations besides $pp$-FSI,
even at high beam energies.

Another important consequence of this work is that we can finally
ascertain with greater confidence the quality of two- and three-nucleon
force models one uses to describe $pd$ observables; 
any disagreement with high quality $pd$ data may now be solely attributed
to the underlying nuclear interaction. In the framework of the present
study we reanalyzed the contribution of $\Delta$-isobar degrees of freedom
to three-body breakup observables.
The largest $\Delta$ effects take place in analyzing powers for given
configurations. Nevertheless the lack of high quality analyzing
power data on a broad spectrum of configurations prevents a full
evaluation of the $\Delta$ effects in $pd$ breakup.
The situation is even worse in three-body photodisintegration of $\He$,
where there are neither kinematically complete experiments without
polarization nor any analyzing power data available.

\begin{acknowledgments}
The authors thank St.~Kistryn and H.~Paetz gen.~Schieck for providing 
experimental data.
A.D. is supported by the FCT grant SFRH/BPD/14801/2003,
A.C.F. in part by the FCT grant POCTI/FNU/37280/2001,
and P.U.S. in part by the DFG grant Sa 247/25.
\end{acknowledgments}

\bibliographystyle{prsty}

\end{document}